\documentclass{article}

\usepackage{graphicx}
\usepackage{amssymb,amsfonts,amsmath}
\usepackage{natbib}

\title{Discrete flow mapping: transport of phase space densities on triangulated surfaces}
\author{David J. Chappell$^{1}$, Gregor Tanner$^{2}$, Dominik L\"ochel$^{3}$ and Niels
S{\o}ndergaard$^{3}$\vspace{2mm}\\
1. School of Science and Technology, Nottingham Trent University, NG11 8NS,  UK.\\ 2. School of Mathematical Sciences, University of
Nottingham, NG7 2RD, UK.\\ 3. inuTech GmbH,  F\"urther Strasse 212,
90429 Nuremberg, Germany}
\date{}
\begin{document}
\maketitle
\abstract{Energy distributions of high frequency linear wave fields
are often modelled in terms of flow or transport equations with ray
dynamics given by a Hamiltonian vector field in phase space.
Applications arise in underwater and room acoustics,
vibro-acoustics, seismology, electromagnetics, and quantum
mechanics. Related flow problems based on general conservation laws
are used, for example, in weather forecasting or molecular dynamics
simulations. Solutions to these flow equations are often large
scale, complex and high-dimensional, leading to formidable
challenges for numerical approximation methods. This paper presents
an efficient and widely applicable method, called {\em discrete flow
mapping}, for solving such problems on triangulated surfaces. An
application in structural dynamics - determining the vibro-acoustic
response of a cast aluminium car body component - is presented.}


\maketitle
\section{Introduction}

Wave energy transport in the high frequency limit can be described
in terms of the underlying ray dynamics, neglecting interference and
other wave effects. The problem is thus reduced to tracking ray
densities in phase space and becomes part of a wider class of mass,
particle or energy transport problems driven by an underlying
deterministic velocity field. Applications are found, for example,
in fluid dynamics (\cite{CCMM04}), weather forecasting
(\cite{SR10}), linear wave dynamics or in general in describing the
evolution of phase space densities of a dynamical system. The
governing equations for the particle or ray densities can often be
written in terms of conservation laws (\cite{LeV92}), an example is
the Liouville equation describing the evolution of ray densities in
phase space for Hamiltonian flows. This class of problems is
particularly interesting in the context of approximating linear wave
equations in terms of their underlying ray dynamics in the short
wavelength limit

Numerical approaches for solving transport problems of this kind are
typically based on the method of characteristics, that is, the
solutions are found along trajectories or rays determined by the
underlying vector field. The flow equations can be formulated in
terms of a linear propagator, the so-called Frobenius-Perron (FP)
operator (see, for example, \cite {Cvi12}), which describes the
evolution of phase space densities in time.  Numerically efficient
methods for solving flow problems in more than one dimension for a
wide class of physically relevant systems are still non-existent. A
variety of techniques have been developed based on a FP-operator
approach, however, all with a fairly limited range of applicability.
Difficulties arise due to the high-dimensionality of the phase space
and the singular nature of the operator describing the underlying
deterministic dynamics. One approach for dealing with such problems
is Ulam's method (see e.g.\ \cite{JD96}), which is based on
subdividing the phase space into distinct cells and considering
transition rates between these phase space regions. Other methods
include wavelet and spectral methods for the infinitesimal
FP-operator (\cite{JK09, FJK11}) and periodic orbit expansion
techniques (\cite{LC10, Cvi12}). The modelling of many-particle
dynamics, such as protein folding, has been approached using short
trajectories of the full, high-dimensional molecular dynamics
simulation to construct reduced Markov models (\cite{Noe09}). For a
discussion of convergence properties of the Ulam method in one and
several dimensions, see \cite{BM01} and \cite{BKL02}, respectively.

More direct methods are based on tracking swarms of trajectories in
phase space often referred to as {\em ray tracing}, see for example
\cite{Cer01}. Methods related to ray tracing but tracking the
time-dynamics of interfaces in phase space, such as moment methods
and level set methods, have been developed by \cite{OF01},
\cite{ER03}, \cite{YC06} and \cite{JB07} amongst others. They find
applications in acoustics, seismology and computer imaging, albeit
restricted to problems with few reflections; for an excellent
overview, see \cite{OR07}. In the following we focus on ray-tracing
approximations of linear wave problems, although the methodology
developed here can be used in a more general context.

Ray tracing and tracking methods often become inefficient when
considering stationary (wave) problems in bounded domains, or in
general for ray tracing problems including multiple scattering
trajectories and chaotic dynamics. An example is the wave field in a
finite cavity driven by a continuous monochromatic excitation. Here,
multiple reflections of the rays and complicated folding patterns of
the associated level-surfaces often lead to an exponential increase
in the number of branches that need to be considered. It is thus
necessary to employ approximation methods for the associated
ray-tracing solutions, often based on ergodicity and mixing
assumptions of the underlying ray dynamics. A popular tool amongst
the mechanical engineering community in the context of
vibro-acoustic modelling is statistical energy analysis (SEA) (see
for example \cite{RL69}, \cite{K92} and \cite{RL95}). A related
method for electromagnetic fields is the random coupling model,
which makes use of random field assumptions  (see \cite{RCM}). In
SEA the structure is subdivided into a set of subsystems and
ergodicity of the underlying ray dynamics as well as
quasi-equilibrium conditions are postulated. The result is that the
density in each subsystem is taken to be approximately constant
leading to greatly simplified equations based only on coupling
constants between subsystems. The disadvantage of these methods is
that the underlying assumptions are often hard to verify {\em a
priori} or are only justified when an additional averaging over
`equivalent' subsystems is considered. The shortcomings of SEA have
been addressed by \cite{RL92}, \cite{RL94} and more recently in a
series of papers by \cite{AL98, AL02, AL06}.

A computational method called {\em dynamical energy analysis} (DEA),
which is based on an FP-operator approach, has been introduced by
\cite{GT09} and further developed in \cite{CGT11}. DEA is an
Ulam-type method subdividing configuration space into smaller
subsystems, but using a basis approximation (spectral method) to
obtain an improved resolution of the phase space density. By
increasing the resolution in both the position and momentum
variable, one systematically interpolates between SEA and full ray
tracing thus relaxing the underlying ergodicity and
quasi-equilibrium assumptions in SEA. A more computationally
efficient approach using a boundary element method for the spatial
approximation has been applied to both two and three dimensional
problems in \cite{CTG12} and \cite{CT13}. A major advantage of DEA
is that by removing the SEA requirements of diffusive wave fields
(equivalent to the ergodicity assumption) and quasi-equilibrium
conditions, the choice of subsystem division is no longer critical.
The resulting increase in flexibility in this choice leads to a much
more widely applicable method.

\begin{figure}
\vspace{-1.5cm} \centering
\includegraphics[width=13cm]{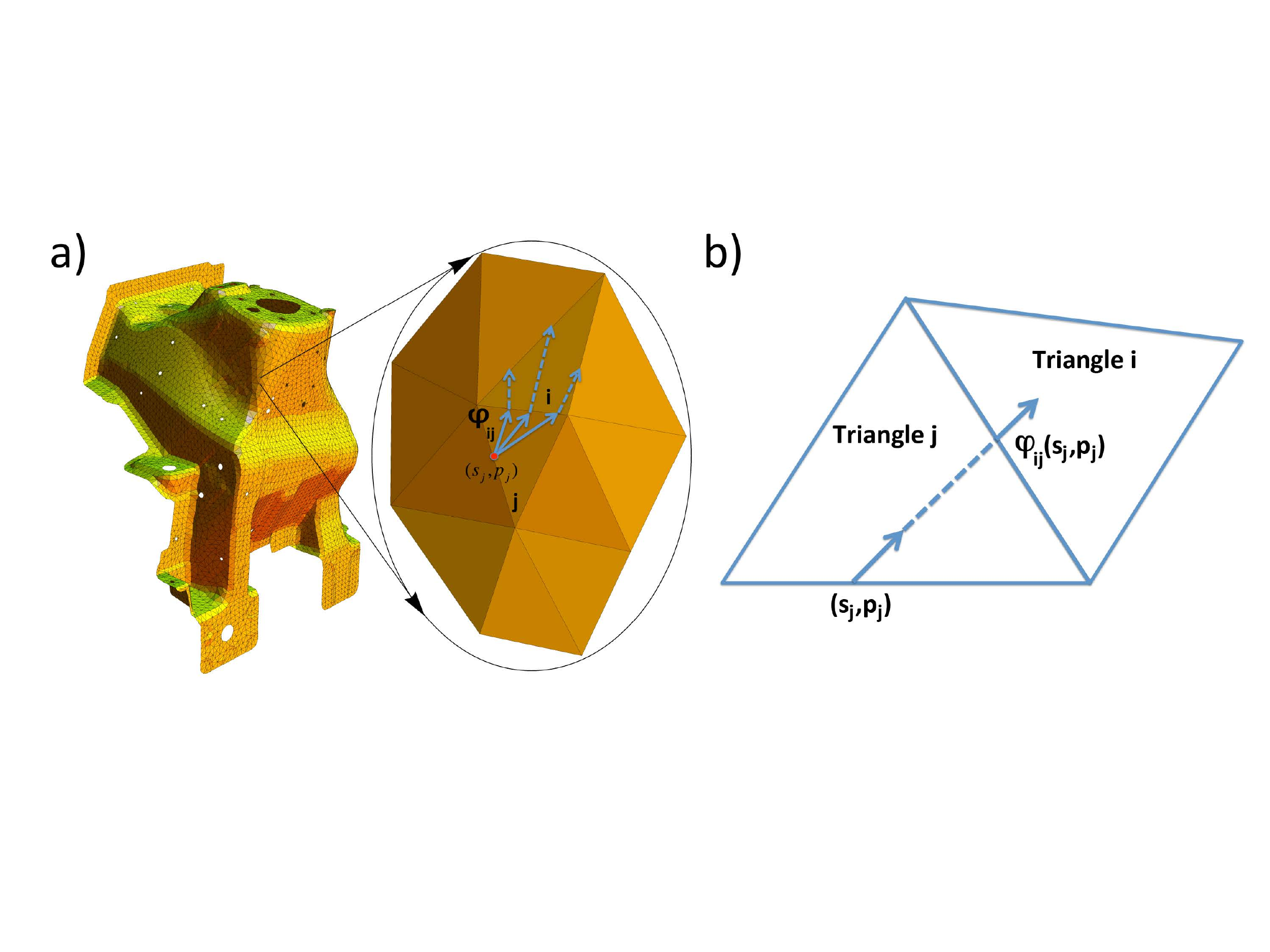}
\vspace{-2.5cm} \label{boundary-map} \caption{(a) Discrete flow
mapping on a mesh of a thin aluminium shell (shock tower) from a
Range Rover; (b) the boundary flow map $\varphi_{ij}$ between a pair
of adjacent triangles. (Online version in colour.)}
\end{figure}

In this work, we exploit the freedom in choosing the subsystems by
extending DEA towards solving the Liouville equation (or other
hyperbolic equations) on meshes. Meshed surfaces are replete in
numerical simulation problems thanks largely to the huge popularity
of finite element methods. Considering the elements of a mesh as our
basic domains therefore automatically renders our techniques
applicable to a wide class of problems including the ability to
handle complex geometries; an example of which is given in Fig.\
\ref{boundary-map}a. A highly efficient solution procedure can be
developed for triangulated surfaces since the problem is simplified
to considering the local ray dynamics in flat planar regions. In
addition, solving phase space flow equations on meshes makes it
possible to consider geodesic dynamics on curved surfaces or ray
dynamics in non-homogeneous media. This opens up the range of
possible applications enormously, for example, in modelling high
frequency vibrations of thin elastic shells (see \cite{AN95} and
\cite{AN96}) or in underwater acoustics (\cite{FJ93}).

The main outcome of this paper will be the introduction and
development of the {\em Discrete Flow Mapping} (DFM) technique, a
new and efficient method for solving stationary phase space flow
equations in complex domains. DFM has similarities to both finite
volume methods and boundary integral methods, combining the
advantages of each. One achieves the reduction in dimensionality to
the boundary of each sub-domain characteristic of boundary integral
methods, which in phase space is a reduction by two. In addition one
can treat non-homogeneous domains as in finite element and finite
volume methods by applying an approximation of local homogeneity and
refining the mesh to improve this approximation. A great advantage
of DFM is that it can be applied directly to existing finite element
models with relatively coarse meshes, and provides a solution for
the high frequency case which automatically contains the geometric
details absent from SEA-type methods. In addition, the flow
directivity can be resolved, unlike in an SEA model, where
ergodicity is assumed.

The paper is structured as follows: we will give an integral
equation formulation for the stationary Liouville equation on
triangulated surfaces in Sec.\ \ref{sec:int-eq}. We then detail the
implementation of DFM in Sec.\ \ref{sec:impl}; using the linearity
of the integral operator, we approximate the solution using a
mixture of boundary element and spectral methods. We also discuss
the treatment of reflection and transmission on interfaces, such as
at edges or due to abrupt changes in material parameters. Finally,
in Sec.\ \ref{sec:num} DFM is applied to model the phase space flow
on a sphere and the geodesic flow on an irregularly shaped car body
part, demonstrating its power and efficiency in practice.

\section{Integral Equation Formulation}
\label{sec:int-eq}

We will focus on triangulated surfaces
$\Omega=\bigcup_{j=1}^{N_{\Omega}}\Omega_{j}\subset\mathbb{R}^3$
consisting of $N_{\Omega}$ triangles $\Omega_j$,
$j=1,...,N_{\Omega}$ such as depicted in Fig.\ \ref{boundary-map}a.
We consider piecewise constant Hamiltonians of the form
$H_j(r,p)=c_j|p|=1$ in $\Omega_j$ describing the energy of a flow,
where $c_j$ is the flow velocity for $r\in\Omega_j$ and the momentum
coordinate $p$ lies on a circle of radius $c_j^{-1}$. This
Hamiltonian is associated to the Helmholtz equation with
inhomogeneous wave velocity $c(r)$ (see \cite{OR07}). In Sec.\
\ref{sec:impl}(\ref{sec:trans}) we also consider vectorial wave
equations, such as plate equations, which include different wave
modes. In this case the wave propagation needs to be characterised
by more than one Hamiltonian per triangle $j$, that is, we need to
consider Hamiltonians $H^l_j = c_j^l |p|$, where $l$ refers to the
mode type. We restrict our discussion to the scalar case in the
following for simplicity of notation.

Let us denote the phase space on the boundary of the triangle
$\Omega_j$ as $Q_j = \partial\Omega_{j}\times(-c_j^{-1},c_j^{-1})$.
The associated coordinates are $X_{j}=(s_j,p_j) \in Q_j$ with $s_j$
parameterising $\partial\Omega_j$, the boundary of the $j$th
triangle, and  $p_j \in (-c_j^{-1},c_j^{-1}) $ parameterising the
component of the inward momentum (or slowness) vector tangential to
$\partial\Omega_j$. We denote the boundary flow map as $
\varphi_{ij}:Q_j\rightarrow Q_i$ which takes a vector in $Q_j$ and
maps it under the flow given through $H_j$ to a vector in $Q_i$, see
Fig.\ \ref{boundary-map}b. Note that $\varphi_{ij}$ is generally
only defined on a subset of $Q_j$, namely the preimage of $Q_i$,
that is $\varphi_{ij}^{-1}(Q_i)\subseteq Q_j$. This preimage is
empty if $\Omega_j$ and $\Omega_i$ are not adjacent (see Fig.\
\ref{boundary-map}).

The stationary density $\rho(X_{i})$ on $Q_i$, $i=1,...,N_{\Omega}$,
due to an initial boundary distribution $\rho^{(0)}$ on $Q_j$,
$j=1,...,N_{\Omega}$, may be computed using the following boundary
integral equation (see \cite{GT09}, \cite{CTG12} and \cite {CT13}),
\begin{equation}\label{GovIE}
(I-\mathcal{B})\rho(X_{i})=\rho^{(0)}(X_{i}),
\end{equation}
where
\begin{equation}\label{BoundaryOperator}
\mathcal{B}\rho(X_{i}):=\sum_j
\int_{-c_j^{-1}}^{c_j^{-1}}\int_{\partial \Omega_j}
K_{\Gamma}(X_{i},X_{j})\rho(X_{j})ds_{j}dp_j,
\end{equation}
and $K_{\Gamma}$ describes the propagation of the flow. Here we
consider purely deterministic flows with
$K_{\Gamma}(X_{i},X_{j})=w(X_{i})\,
\delta(X_{i}-\varphi_{ij}(X_{j}))$. A diffusion component may be
added by replacing the $\delta$-distribution with a finite-width
kernel. In general, the weight function $w$ contains
reflection/transmission probabilities at boundaries and a
dissipative term of the form $\exp(-\mu L)$, where $L$ is the length
of the flow trajectory and $\mu$ is a damping coefficient. For the
case $w=1$, the transfer operator $\mathcal{B}$ is of
Frobenius-Perron type with a maximum eigenvalue one. In this case
the problem, Eq.\ (\ref{GovIE}), becomes ill-posed and thus we
consider the case $\mu>0$ henceforth. One can also obtain a
well-posed problem using other forms of dissipation, such as an
absorbing or open boundary region (\cite{CTG12}). Note that since
the flow map only maps to neighbouring triangles, a matrix
representation of $\mathcal{B}$ over the whole of
$\bigcup_{i=1}^{N_{\Omega}}Q_{i}$ is in general sparse. Equation
(\ref{GovIE}) is a boundary integral formulation for the stationary
Liouville equation as detailed in \cite{CT13}. The derivation of
$\rho^{(0)}$ for a high frequency point source is also given in
\cite{CT13}.

\section{Implementation of discrete flow mapping}
\label{sec:impl}
\subsection{Discretisation}
Typically it is assumed that $\Omega$ consists of a large number of
triangles $N_{\Omega}$ describing a geometrically complex domain. A
discrete approximation of $\rho$ is sought in phase space
coordinates on the triangle boundaries. The spatial approximation is
given by taking polynomial basis functions on each triangle edge
$\alpha$ with $\alpha =1,2,3$. That is, we split the approximation
at corners, see \cite{CGT11} for further discussion on the benefits
of this choice. In the following, we use piecewise constant
functions on the triangle edges for simplicity. In contrast to the
position coordinate, the momentum coordinate has support on the
interval $(-c_j^{-1},c_j^{-1})$ only. It is
therefore proposed to employ a Legendre polynomial basis
approximation in this coordinate (see \cite{CTG12}).
A key advantage of these choices is that the integrand in the operator (\ref{BoundaryOperator})
remains a very simple function of the position argument and the corresponding integral can be
performed analytically. This dramatically reduces the costs of evaluating (\ref{BoundaryOperator})
compared to the implementation used in \cite{CTG12}. 

\begin{figure}
\centering
\includegraphics[width=10cm]{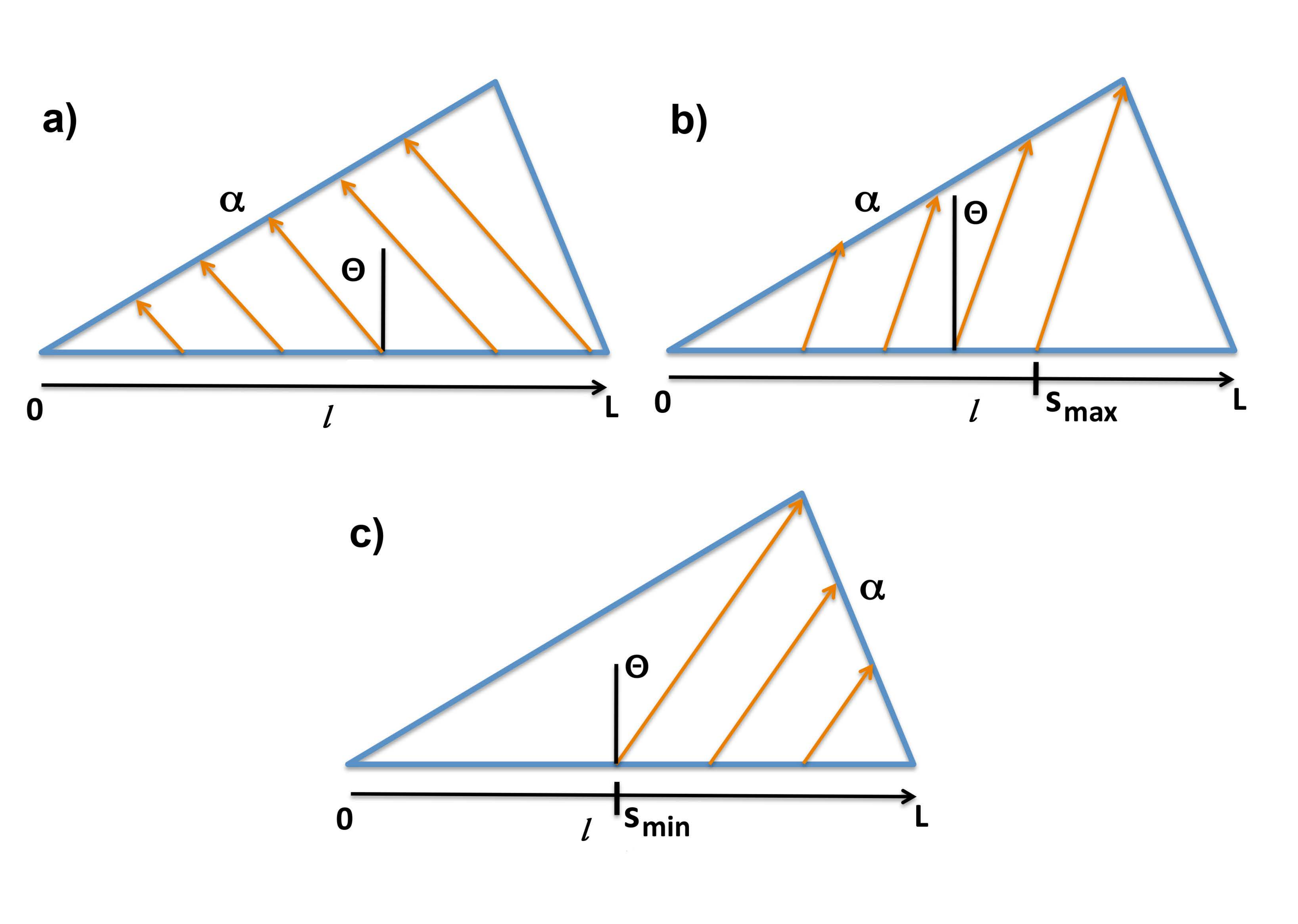}
\vspace{-.5cm} \caption{Admissible ranges $(s_{min},s_{max})$ for
$s_j$ when a ray travels from edge $l$ of triangle $j$ with fixed
direction coordinate $p_j=\sin(\theta)/c_j$ to an opposite edge
$\alpha$: (a) $s_j$ can take all values in $(0,L)$ on the edge $l$
(i.e. $s_{min}=0$, $s_{max}=L$); (b) $s_j\in (0,s_{max})$; (c)
$s_j\in (s_{min},L)$. (Online version in colour.)} \label{sminmax}
\end{figure}

The overall approximation on $Q_i$ for $i=1,...,N_{\Omega}$ is then
of the form
\begin{equation}
\rho^{\Gamma}(X_{i})\approx\sum_{\alpha=1}^{3}\sum_{\beta=0}^{N_{p}}\rho_{(i,\alpha,\beta)}b_{\alpha}(s_i)\tilde{P}_{\beta}(p_{i}),
\end{equation}
where $N_{p}$ is the order of the momentum basis expansion. The
momentum basis functions are given by
\begin{equation}\label{2DLeg}
\tilde{P}_{\beta}(p_i)={\sqrt{c_i}}P_{\beta}(c_{i}p_{i}),
\end{equation}
where $P_{\beta}$ is the Legendre polynomial of order $\beta$. The
piecewise constant spatial basis functions are given by
$b_\alpha(s_i)=2^{-1/2}/\sqrt{A_{\alpha}}$ for $s_i$ on the edge
$\alpha$, and zero elsewhere. Here $A_{\alpha}$ is the length of the
edge $\alpha\in\{1,2,3\}$ of $\Omega_{i}$. Imposing a weak form of
the integral equation (\ref{GovIE}) using the orthonormal inner
product for Legendre polynomials $<\cdot,\cdot>$ yields
\begin{equation}\label{Neum-ser-disc}
(I-B)\underline{\rho}=\underline{\rho}^{0},
\end{equation}
where $I$ is the identity matrix,
\begin{eqnarray}\label{op_approx0}
\rho_{(i,\alpha,\beta)}^{0} & = & \:\left<\rho^{(0)}(s_i, p_i),
b_{\alpha}(s_i)\tilde{P}_{\beta}(p_{i})\right>,\\ \nonumber
B_{(i,\alpha,\beta),(j,l,m)} & = &
\left<\mathcal{B}\left(b_{l}(s_i)\tilde{P}_{m}(p_{i})\right),
b_{\alpha}(s_i)\tilde{P}_{\beta}(p_{i})\right>,
\end{eqnarray}
and the vectors $\underline{\rho}$, $\underline{\rho}^{0}$ have
entries given by $\rho_{(i,\alpha,\beta)}$,
$\rho_{(i,\alpha,\beta)}^{0}$, respectively. Expanding the inner
product in (\ref{op_approx0}) and using the definition of the
transfer operator $\mathcal{B}$ (\ref{BoundaryOperator}), the
discretised transfer operator $B$ acting on $Q_{i}$ for all
$i=1,...,N_{\Omega}$ may be expressed as
\begin{eqnarray}\label{op_approx}
&&{\displaystyle B_{(i,\alpha,\beta),(j,l,m)}} \\ \nonumber
 &&{\displaystyle=\frac{2m+1}{4}\int_{Q_i}\int_{Q_j}\tilde{P}_{\beta}(p_i)b_{\alpha}(s_i)K_{\Gamma}(X_i,X_j)\tilde{P}_{m}(p_j)b_{l}(s_j)ds_{j}dp_j ds_{i}dp_i }\vspace{2mm}\\ \nonumber
&& {\displaystyle=\frac{2m+1}{4}\int_{Q_j}
w(\psi_{ij}(X_j))\tilde{P}_{\beta}(\varphi^{p}_{ij}(X_{j}))b_{\alpha}(\varphi^{s}_{ij}(X_j))\tilde{P}_{m}(p_j)b_{l}(s_j)ds_{j}dp_j.}
\end{eqnarray}
Here we write $\varphi_{ij}=(\varphi^{s}_{ij},\varphi^{p}_{ij})$ to
denote the splitting of the position and momentum parts of the
boundary map. Using the properties of the weight function and the
spatial basis functions, Eq.\ (\ref{op_approx}) simplifies to
\begin{eqnarray}\label{op_approx2}
&&B_{(i,\alpha,\beta),(j,l,m)} = \\ \nonumber
&&\frac{2m+1}{8\sqrt{A_\alpha A_l }}\int_{-c_j^{-1}}^{c_j^{-1}}
\hspace{-3mm}
\tilde{P}_{m}(p_j)\int_{s_{min}(p_j,\alpha,l)}^{s_{max}(p_j,\alpha,l)}
\hspace{-6mm} w_{rt}(\varphi^{p}_{ij}(X_j))e^{-\mu
L_{ij}(s_j,\varphi^s_{ij}(X_j))}\tilde{P}_{\beta}(\varphi^{p}_{ij}(X_{j}))ds_{j}dp_j.
\end{eqnarray}
Here, $w_{rt}$ is equal to the transmission probability $w_t$ when
$i\neq j$, and is equal to the reflection probability $w_r=1-w_t$
otherwise. Higher dimensional transmission/reflection matrices arise
in the case of multi-mode dynamics as discussed in the next section.
Also $L_{ij}$ is the length of the trajectory from the point
represented by $s_j$ to $\varphi^s_{ij}(X_j)$, and
$(s_{min},s_{max})$ is the admissible range of values for $s_j$.
Restriction to this range is necessary to ensure that a ray starting
on edge $l$ of triangle $j$ with direction coordinate $p_j$ will be
transferred to a particular adjacent edge $\alpha$ of triangle $i$
as shown in Fig. \ref{sminmax}. Note that for flat polygonal
elements such as triangles, $\varphi^{p}_{ij}(X_j)$ in fact only
depends on $p_j$, and hence only the damping term in equation
(\ref{op_approx2}) retains dependence on $s_j$. The inner integral
thus has a simple form and can be computed analytically, leaving
only a single integral to be evaluated numerically. It is this step
which leads to vast improvements in the computing times, and enables
us to consider many thousands of elements with high order
approximations in direction space. Two key points are:
\begin{itemize}
\item Triangulation is used to ensure that the elements of a piecewise linear mesh are flat. This enables us to deal with complex enclosures/surfaces whilst keeping the system locally simple.
\item  Due to the semi-analytic phase space integration, working with many thousands of subsystems and high order direction space approximations becomes tractable.
\end{itemize}
It is worth pointing out that the analytic integration method is not
restricted to triangular elements, but works for any flat polygons.
The general idea of the DFM is related to the work in \cite{CGT11}
and \cite{CT13}. However, in these papers, the DEA method was
developed for general subsystems of arbitrary shape and the double
integrals in the equations equivalent to Eq.\ (\ref{op_approx2})
needed to be computed entirely numerically. Computing the matrix $B$
is then time consuming and the computations were restricted to a
small number of subsystems (typically $10$ or less).

\subsection{Transmission at interfaces and ray tracing on curved surfaces}
\label{sec:trans}
\begin{figure}
\centering
\includegraphics[width=5.5cm]{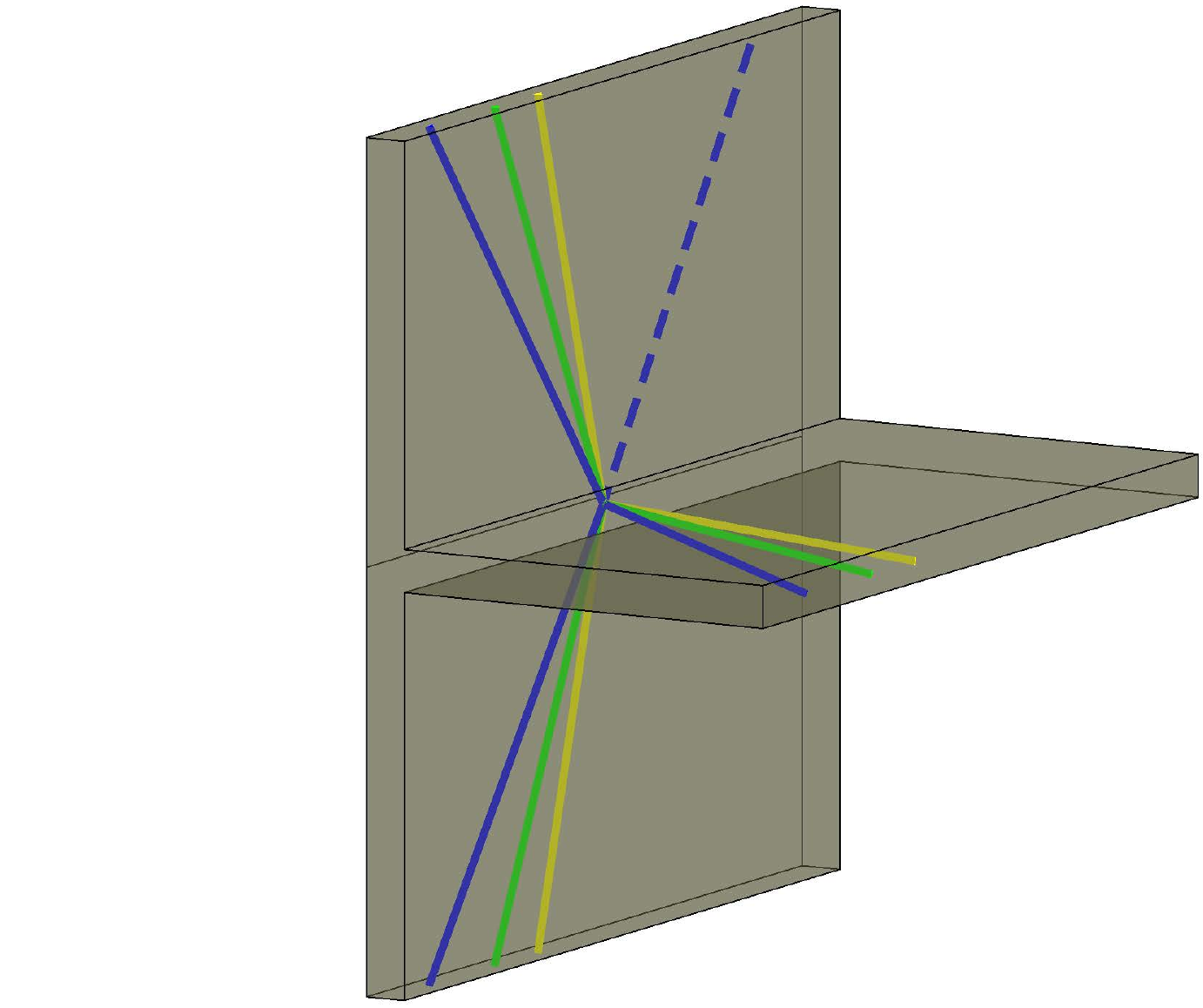}
\caption{T-joint as an example of a junction between 3 plates. The
dashed line indicates an incoming wave, the solid lines represent
outgoing bending, shear and pressure waves. (Online version in
colour.)} \label{fig:joint}
\end{figure}

For modelling high frequency wave propagation it will be necessary
to take into account transmission (and consequently reflection) at
interfaces with abruptly changing material parameters or due to
edges and branch lines. The latter will arise, for example, when
plates are connected along a common junction such as depicted in
Fig.\ \ref{fig:joint}. These effects are  included in our approach
through transmission/reflection probabilities at triangle boundaries
coinciding with the interface. The reflection/transmission
coefficients are obtained from local wave solutions at the
interface, incorporating the dependence on the direction of the
incoming and scattered waves. A number of scenarios important in the
context of both wave propagation and ray tracing are outlined below.

\paragraph{Scalar wave equations:}
For scalar wave equations and simple planar interfaces, the
transmission coefficients are derived from the wave equation using
continuity of the wave function and its normal derivative for an incident plane
wave (see \cite{JD92}). One thus obtains
\begin{equation}\label{rt-coeff}
w_{t}(k_{i},k_{j},\theta_{j})=\frac{4(k_{i}/k_{j})\cos(\theta_{j})
\cos(\theta_{i})}{((k_{i}/k_{j})\cos(\theta_{j})+\cos(\theta_{i}))^{2}}.
\end{equation}
Here, $k_{j}$ and $k_{i}$ are the wave numbers in the elements
containing the incoming and outgoing rays, respectively, with $k_i =
\omega/c_i$ and $\omega$ is the angular frequency. A change
of wave vector at the interface may be due to  a change in the
material parameters for example. The angle of incidence $\theta_{j}$
of the incoming ray with respect to the normal at the element
boundary is simply $\arcsin(c_j p_j)$. The direction of the outgoing
ray $\theta_{i}$ is determined using Snell's law.

\paragraph{Elastic waves in coupled plates:}
A more general case is represented by interfaces forming junctions
between plates of varying thickness and meeting at arbitrary angles.
The transmission/reflection coefficients for a set of plates coupled
at a common interface (such as depicted in Fig.\ \ref{fig:joint})
may be computed using the methods presented in \cite{LH90} and
\cite{Cra04}. In particular, we consider the connection between
plates as line junctions, that is, the interior properties of the
junction are not modelled and the mass and moment of inertia are
neglected. Let us consider a line junction which couples $n$
different (for simplicity) semi-infinite plates. The boundary
conditions at the line junction correspond to dynamic conditions
involving stresses, moments and kinematic conditions for the
displacement and rotation of the $n$ plates. To construct the
transmission coefficients, we calculate the response of the system
with respect to excitation by an incoming plane wave. The incoming
wave has a fixed wavenumber and a characteristic mode, that is, it
is of bending ($b$), pressure ($p$) or shear ($s$) type. The
outgoing waves typically have components in all $n$ plates and are a
mixture of all mode types. Evanescent modes may be included to
complete the description. Possible material differences between the
plates can lead to different wavenumbers in different plates. For a
given forcing with a particular incoming mode in a particular plate,
we can solve for the unknown modal coefficients in all plates. In
practice, we find the transmission probabilities directly by
calculating the ratio of outgoing to incoming normal power fluxes. A
detailed description can be found in \cite{Tan13}.

\paragraph{Curved surfaces:}
In the case of a geodesic flow on a smooth curved surface it is
necessary to mimic the geodesic paths on the corresponding
triangulated surface. We employ the theory outlined in \cite{JS98}
and \cite{DM05}, making use of the fact that for ray paths not
intersecting vertices on triangulated surfaces, the notions of
shortest and straightest (discrete) geodesic are equivalent (see
\cite{DM05}). Hence the straightest geodesic choice of
$\theta_{j}=\theta_{i}$ will approximate the direction of the
geodesic flow on homogeneous regions of the triangulated surface,
and abrupt changes of surface thickness or material will result in a
Snell Law effect (see \cite{JS98}). A suitable choice of quadrature
method here ensures that the rays never pass through vertices of the
triangulation (although a point source may lie on a vertex), for
example Gaussian quadrature rules where endpoints are never used as
abscissae.

\paragraph{Curved shells - curvature corrections:}
For modelling the vibration of thin shells we need to consider ray
tracing on curved surfaces, where the dynamics are derived from thin
shell theory (\cite{AN95}). In a lowest order approximation, curved
rays again follow the geodesics of the surface (\cite{AN96,GT07}).
This approximation derives from applying thin shell theory in the
high frequency limit as in \cite{AN94} and  \cite{AN95}, which is
valid for wavelengths shorter than the radii of curvature of the
shell, but larger than the thickness. Corrections to the geodesic
ray approximations need to be considered if the local radii of
curvature are of the same order as the wavelength. It is possible to
construct modified ray paths from the dispersion curves given by
thin shell theory (see \cite{AN94}), but this requires a detailed
knowledge of the local curvature. In the interests of keeping the
model as simple as possible we will follow a different approach
here. We treat the meshed structure as a set of plate-like elements
and estimate reflection/transmission properties due to the finite
angle between mesh elements using the plate-junction theory sketched
above and detailed in \cite{LH90} and \cite{Cra04}. This enables us
to mimic wave barriers due to regions
of high curvature as demonstrated in the next section.\\

In each of the cases considered above,
reflection/transmission coefficients
are incorporated as part of the weight function $w$ in
the boundary integral kernel $K_{\Gamma}$ in Eq.
(\ref{BoundaryOperator}), or its finite dimensional approximation.
We assume that the transmission coefficients depend only on the
incoming momentum $p_s$ via the angle of incidence; the outgoing
angle is given by Snell's law taking into account refraction due to
differences in the wave speed across the interface.

\section{Numerical examples}
\label{sec:num} We will demonstrate the efficiency and flexibility
of DFM with the help of two examples. Firstly, we determine the ray
density produced by a point source on a sphere, where an analytic
solution is available for verification. Secondly, we compute the
energy density distribution on the curved surface of a cast
aluminium Range Rover body part and compare with solutions of the
corresponding wave equation obtained using the finite element method
(FEM).

\subsection{Flow on a sphere}

\begin{figure}
\centering
\includegraphics[width=13cm]{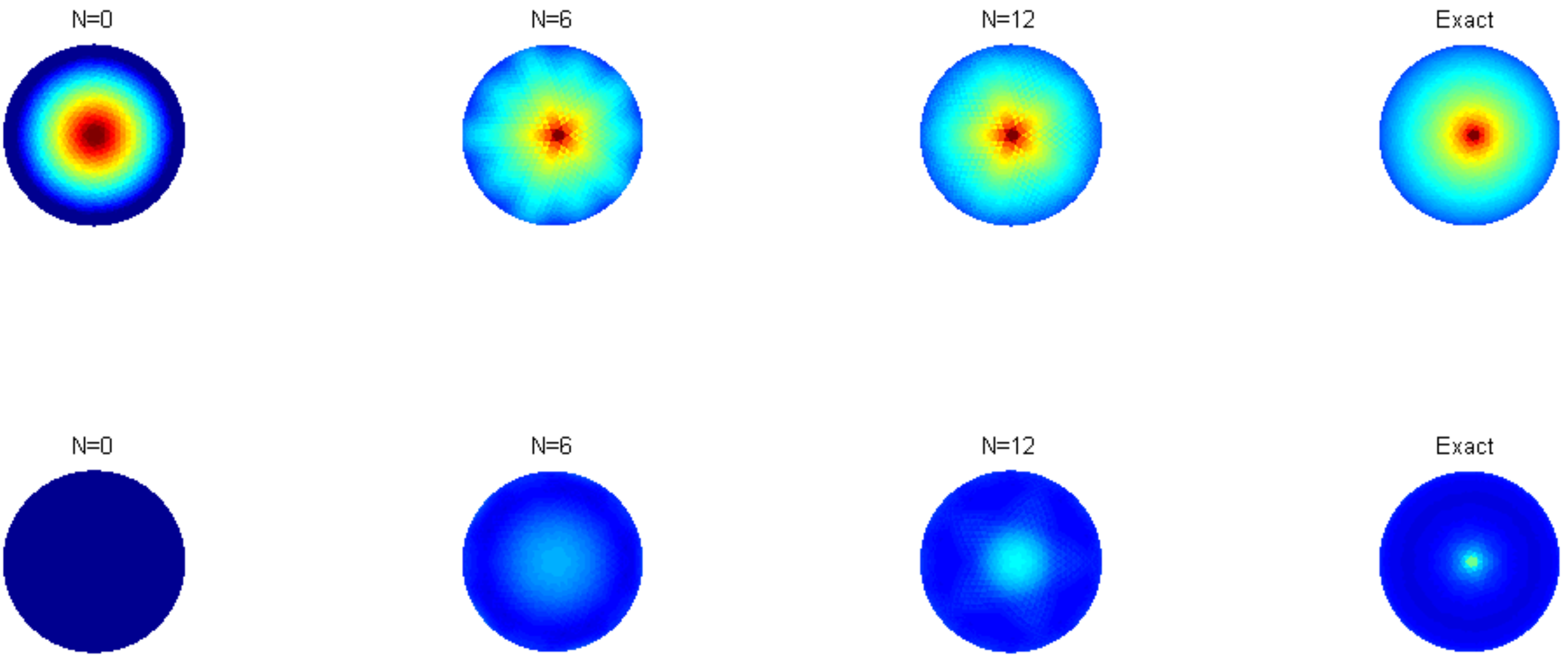}
\caption{Ray density (logarithmic scale) on a sphere computed using
discrete flow mapping showing the source point (upper) and antipode
(lower) for different orders of direction space basis ($N_p$) on a
triangulated sphere with 5120 triangles. The plots on the right show
a representation of the exact ray density over the same
triangulation. (Online version in colour.)} \label{spheremulti}
\end{figure}

The ray density $\rho$ generated by a geodesic flow emanating
continuously from a point source on a sphere can be determined
analytically. As a function of the polar angle $\phi$, with source
point $\phi=0$, the ray density is given by:
\begin{equation}\label{ExactSphere}
\rho(\phi)=\frac{C e^{-\mu \phi}}{(1-e^{-2\pi\mu })\sin(\phi)}.
\end{equation}
Here, $\mu$ is the damping coefficient introduced in Sec.\
\ref{sec:int-eq} and $C$ is a constant depending of the strength of
the source. The derivation of equation (\ref{ExactSphere}) follows
simply from the fact that the ray density on the sphere is inversely
proportional to the element of surface area. The exponential terms
result from damping contributions of the form $\exp(-\mu L)$;
summing over the trajectory lengths $L$ at any point on the sphere
results in a geometric series with a contribution each time the ray
orbits the sphere.

The sphere is thus a good candidate for verifying our approach on an
approximately spherical triangulated surface. The dynamics on the
sphere are integrable and the exact solution for the density
contains a singularity at both poles ($\phi = 0$ and $\pi$) as the
rays are unidirectional along great circles passing through the
source point. This example is therefore slightly atypical since
ergodic or mixing ray dynamics in complex geometries generally lead
to more smoothly distributed ray densities. The sphere is thus a
true challenge for DFM, which due to the finite
order basis approximation will always incorporate diffusive
behaviour and consequently a smoothing of singularities.

Fig.\ \ref{spheremulti}  demonstrates that DFM can deal even with
such a singular case. Higher order implementations of the basis
approximation in direction space lead to an improved resolution of
the refocussed singularity. Table \ref{errorsphere} gives the mean
relative error averaged over the upper hemisphere containing the
source point shown in the upper row of Fig.\ \ref{spheremulti}. The
results are given for Delaunay triangulations with differing numbers
of elements and for different orders of direction space basis. The
results are computed at the centroid of each triangle and compared
against the exact solution for the same value of $\phi$. For the
results presented here we have taken $\mu=1$ and
$C=(800\pi^2)^{-1}$, which is the scaling for a unit excitation of
the Helmholtz equation with $k=100\pi$. The factor is derived from
the high frequency asymptotics of the 2D fundamental solution for
the Helmholtz equation (see \cite{CT13}) and matching the
asymptotics with equation (\ref{ExactSphere}) as the distance from
the source becomes small.

\begin{table}\caption{Mean relative errors computed at triangle centroids and averaged over the upper hemisphere of a Delaunay triangulated sphere with $N_{\Omega}$ triangles and a direction space Legendre polynomial basis approximation of order $N_p$.}\label{errorsphere} 
\centering
 \begin{tabular}{|ccc|}\hline
  {$N_{\Omega}$} & {$N_p$} & {Mean Relative Error }\\
  \hline
  {320} & {4} & {0.1606}\\
  {320}  & {6} & {0.1129}\\
  {320}  & {8} & {0.1142}\\
  \hline
  {1280}  & {8} & {0.08704}\\
  {1280}  & {10} & {0.08884}\\
  \hline
  {5120}  & {10} & {0.06648} \\
  {5120}  & {12} & {0.06212}\\
  {5120}  & {14} & {0.06100}\\
   \hline
  {20480}  & {14} & {0.05116} \\
  {20480}  & {16} & {0.05012}\\ \hline
 \end{tabular}
 \end{table}

\subsection{An Application in Vibro-acoustics}
\label{sec:rangerover} In this section we consider the transport of
high frequency flexural or bending wave energy through curved
structures using thin shell and high frequency ray models. At
present there is wide interest in modelling aluminium shells in the
automotive industry due to a desire for lighter, and hence lower
emission vehicles. In particular, large molded aluminium components
are replacing more traditional multi-component beam-plate
constructions, which has the additional advantage of eliminating
problems due to fatigue at joints. High frequency vibro-acoustic
models based on an SEA treatment will be unsuitable in these
circumstances, since complex geometrical features are not included
in SEA. In addition, a subdivision of the model into subsystems is
not clear cut for such castings as all components are well
connected, see for example the structure in Fig.\
\ref{boundary-map}a representing an aluminium shock tower from a
Range Rover.

\begin{figure*}
\centering
\includegraphics[width=13cm]{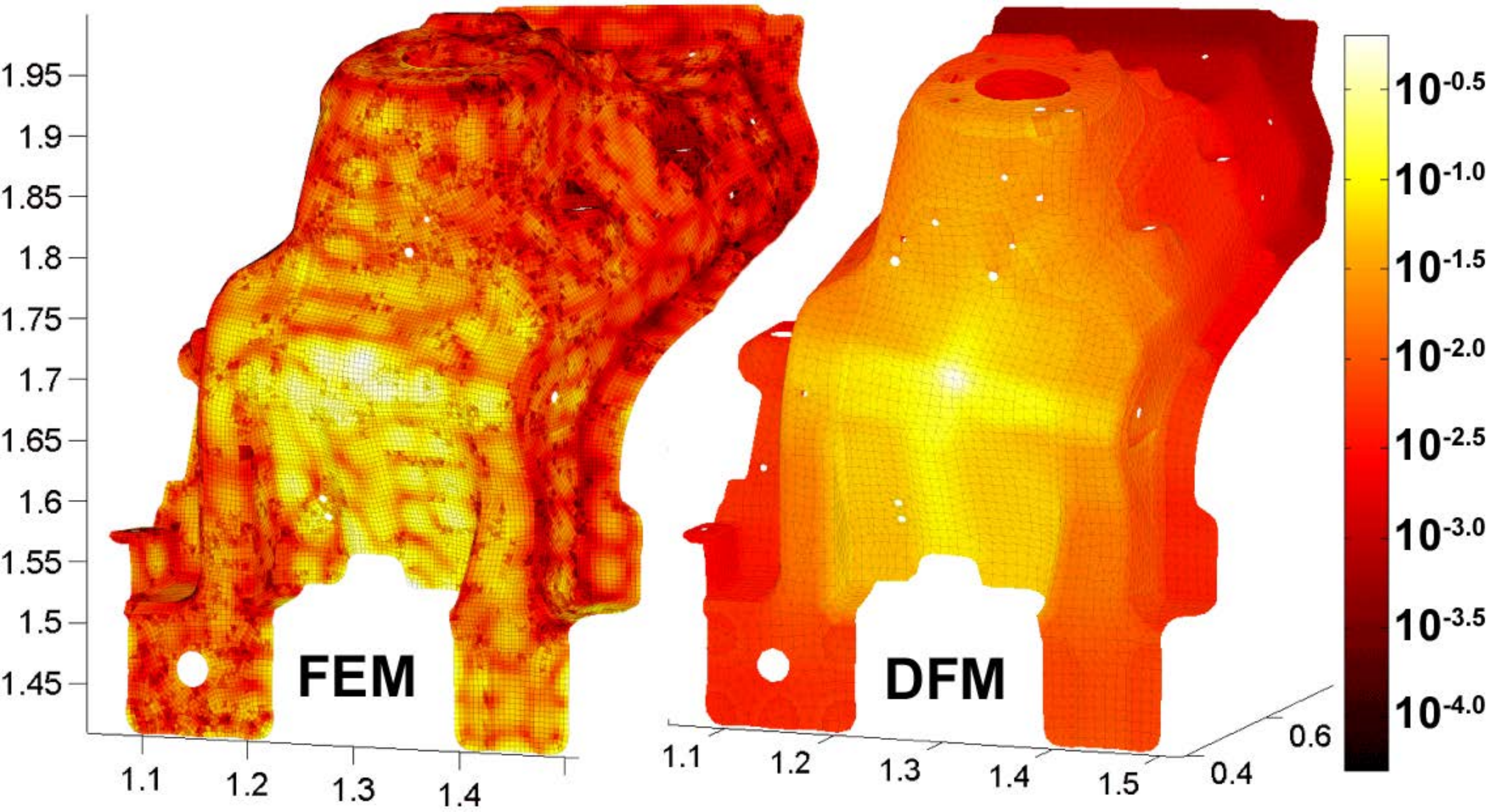}
\caption{Kinetic energy density on a thin aluminium shell estimated
using an averaged full wave finite element model (left) and discrete
flow mapping (right) for $3\%$ hysteretic damping. (Online version
in colour.)} \label{shocktowerstddamp}
\end{figure*}

Discrete flow mapping can overcome these problems, since it can be
easily applied in the framework of existing grids for finite element
models, requires no choice of subsystem division and incorporates
the full geometry and directionality of the energy flow. The
response of a thin molded aluminium car component (Range Rover shock
tower) to a point force applied normally to the surface is modelled
and compared with a FEM approximation for the full wave model
performed using Nastran. Figure \ref{boundary-map}a shows the
problem setup including the mesh and indicating the local DFM map
$\varphi_{ij}$. In order to maintain a tractable model size for the
finite element calculation and to study frequency ranges of
industrial interest, the computation is performed at frequencies
between 8\,kHz and 10\,kHz. This approximately corresponds to a
third of an octave band centred at 9\,kHz. A key assumption of the
thin shell theory is that the radius of curvature is large compared
to the wavelength (see \cite{AN94}), which  is only partly satisfied
for this structure in the frequency band considered. The wavelength
is typically around $3$ to $6\:$cm depending on the shell thickness.
We therefore employ a modified geodesic ray tracing technique by
incorporating reflection/transmission due to finite angles between
mesh elements as described in Sec.\ \ref{sec:impl}(\ref{sec:trans}).
This leads to geodesic trajectories without reflections in
relatively flat areas, but incorporates wave barrier behaviour in
regions of high curvature.

Fig.\ \ref{shocktowerstddamp} shows the DFM result (bending mode
only) compared with the Nastran solution for the kinetic energy
averaged over 41 evenly spaced frequencies spanning the prescribed
range and with a hysteretic damping level of $3\%$, which is typical
for such a structure. The point source is positioned at the front of
the structure. The ray computation is performed using a triangulated
surface consisting of $11\,623$ triangles and with a 6th order
Legendre polynomial basis in direction space. The Nastran grid
contains $40\,670$ elements comprising a mix of piecewise linear
triangles and quadrilaterals. As would be expected one sees more
oscillation in the full wave model. The DFM prediction of the
overall energy flow in regions of both high and low curvature
matches well with the Nastran results. The flow of energy along the
side walls of the central mount, as well as features such as shadow
regions due to holes in the structure and channelling effects due to
variations in the shell thickness can be observed as common in both
plots. Such geometrically dependent features would be entirely
absent from SEA-type models and represent a major advance for high
frequency vibro-acoustic simulation methods.

\begin{figure*}\vspace{-8mm}
 \centering
\includegraphics[width=13cm]{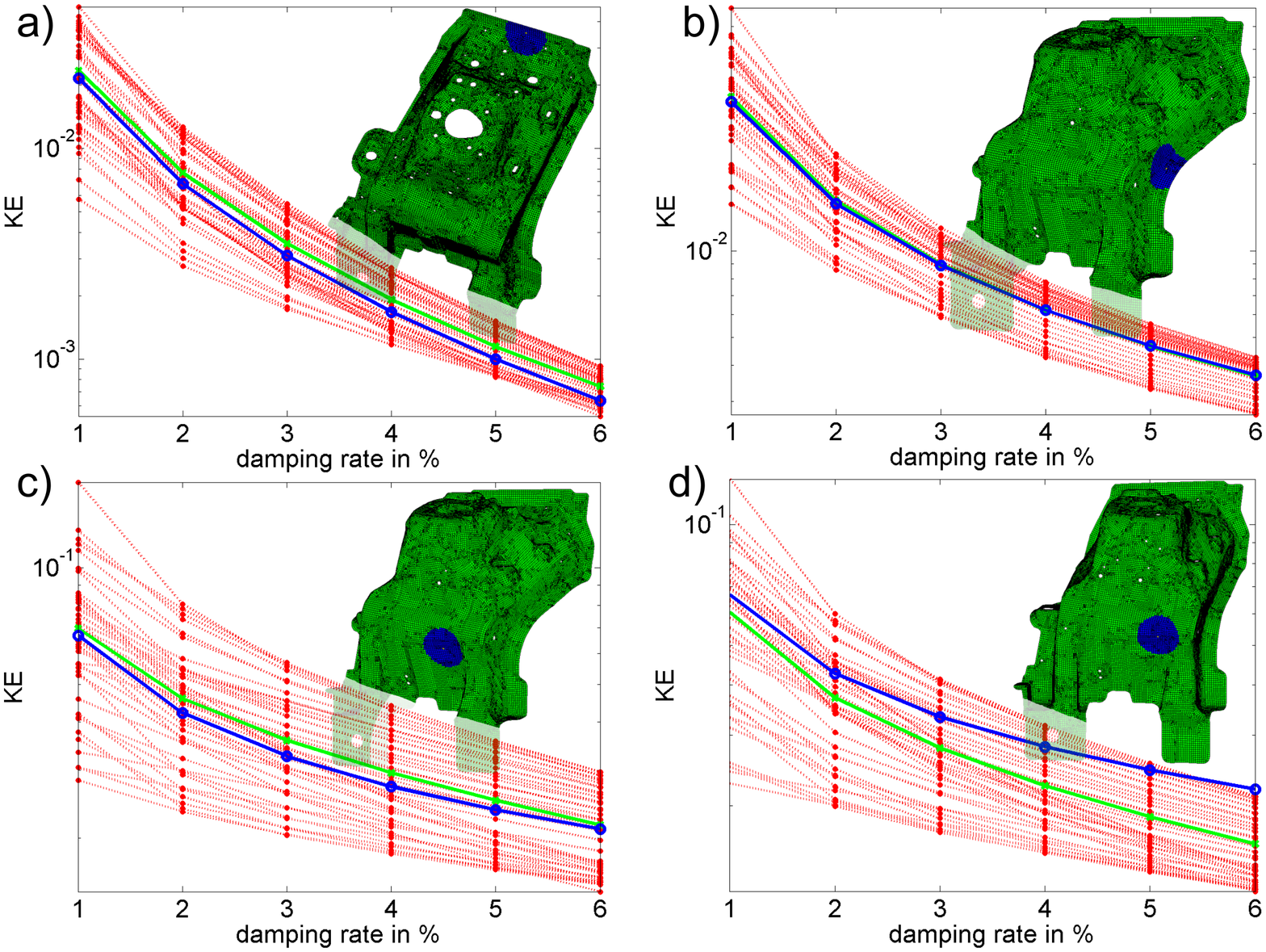}
\caption{Kinetic energy (KE) response against hysteretic damping
level. The KE is averaged over the blue circular receiver regions.
DFM: solid blue line with `o' markers; FEM: dotted red lines with
`$\cdot$' markers; mean FEM: solid green line with `x' markers.
(Online version in colour.)} \label{shocktowerrcpts}
\end{figure*}

Figure \ref{shocktowerrcpts} gives the kinetic energy averaged over
four different receiver regions, which are displayed as the blue
regions on the inset structure plots. The response is now shown for
a range of damping levels between 1$\%$ and 6$\%$, and for receiver
regions with differing levels of proximity to the source point. The
correspondence between the mean of the 41 Nastran solutions and DFM
is remarkably good in Fig.\ref{shocktowerrcpts} (a) to (c). The
regions are spread across the whole structure and provide strong
verification of our approach. In Fig.\ \ref{shocktowerrcpts} (d),
the DFM results deviate slightly for high damping. Note that this
region is close to the region in Fig.\ \ref{shocktowerrcpts} (c) and
the deviations thus reflect local variations in the FEM solution.
The results clearly demonstrate that curvature related wave-barrier
effects are correctly predicted by DFM using local
reflection/transmission matrices.

The computations in this section were performed in parallel on four
cores of a desktop PC in less than 6 minutes. The code was written
in C++ with openMP. The large sparse non-symmetric linear system of
dimension 732\,249 which arises has been solved efficiently and
accurately using a stabilized bi-conjugate gradient iterative
solver.

\section{CONCLUSIONS}
We present {\em discrete flow mapping} as a highly efficient method
for solving stationary phase space flow equations in complex domains
and apply the method to two illustrative examples. In particular we
highlight the versatility of the method, its capability to handle
large scale models efficiently and emphasise the fact that both
geometric details and flow directivity are fully included in the
calculation at a moderate computational cost. In the special case of
ray focusing, the method clearly captures the foci albeit not
reproducing the actual singularity. In the vibro-acoustic example,
the full complexity of the model is taken into account via the mesh
functionality of DFM. Deviations from a pure geodesic flow due to
regions of high curvature have been reproduced without compromising
the efficiency of the method. DFM can thus serve as a practical tool
for solving phase space transport problems with applications in
engineering ranging from acoustics to structural mechanics.
Extensions of the method to electrical engineering, as well as to
fluid mechanics and other flow problems are evident.

\section*{Acknowledgement}
Support from the EPSRC (grant EP/F069391/1 and a follow-on {\em
Knowledge Transfer} grant) and the EU (FP7 IAPP grant MIDEA) is
gratefully acknowledged. The authors also gratefully acknowledge the
support, guidance, data and Nastran access provided by Stephen
Fisher and Jaguar Land Rover.

\end{document}